# An atomic force microscopy and optical microscopy study of various shaped void formation and reduction in 3C-SiC films grown on Si using chemical vapor deposition

A. Gupta, J. Sengupta, C. Jacob *

Materials Science Centre, Indian Institute of Technology, Kharagpur, 721302, India

## Abstract

The formation of various uncommon shaped voids along with regular triangular and square voids in the epitaxial 3C-SiC films on Si has been investigated by optical microscopy and atomic force microscopy. Heteroepitaxial growth of 3C-SiC films on Si (001) and (111) substrates has been performed using hexamethyldisilane in a resistance-heated chemical vapor deposition reactor. The influence of the orientation of the Si substrate in determining the shape of the voids has clearly been observed. In addition, the growth period and the growth-temperature have been considered as the major parameters to control the size, density and shape of the voids. Generally, voids are faceted along {111} planes, but depending upon growth conditions, other facets with higher surface energy have also been observed. Finally the size and density of the voids are remarkably reduced, by suitable growth technique.

## 1. Introduction

3C-SiC is an excellent semiconducting material for high temperature, high power and high frequency electronic devices due to its wide bandgap ($\sim 2.23$ eV), high electron mobility, high saturated electron drift velocity and high thermal stability, etc. [1]. In addition, amorphous and polycrystalline SiC are often used as a coating material for structural applications and for micro-electro-mechanical systems applications due to their excellent mechanical properties and chemical inertness. Chemical vapor deposition (CVD) is most commonly utilized for the epitaxial growth of 3C-SiC on Si. Since 1983, the heteroepitaxial growth of 3C-SiC on Si substrates has become very attractive, but due to the large mismatch of the lattice constant (20%) and of the thermal expansion coefficient (8%) between 3C-SiC and Si, a high density of defects such as dislocations, stacking faults, microtwins etc. is found at the interface, which drastically influences the grown films [2]. These defects can be reduced by using 2°–4° tilted Si substrates and preparing a buffer layer on the Si substrate by carbonization at high temperature before the growth [3]. During this carbonization process, voids are generated at the SiC/Si interface. Voids, which are the common interfacial defect associated with heteroepitaxial growth of 3C-SiC on Si, are created by the coalescence of Si vacancies, which result from Si out-diffusion from the Si substrate. Growth of voids is most rapid before and during the coalescence of SiC nuclei. For the most part, voids seem to remain as empty sites during their expansion, with the grown SiC layer bridging the voids. However, there are also reports of SiC material protruding from SiC layer bottom surfaces [4,5] or of voids partly filled with SiC [6–8] indicating a certain ingrowth of SiC into voids or the substrate, respectively. The formation of voids generates microstructural defects and roughness at the interface. As a result, it poses a problem in SiC/Si device applications, such as p-n junction diodes, metal oxide semiconductor field effect transistor (MOS-FET) and heterojunction bipolar transistor. The key issue for obtaining a power-switching MOS-FET using 3C-SiC is to reduce the leakage-current of the p-n junction by eliminating planar defects like stacking faults [9]. Reduction of planar defects in 3C-SiC through heteroepitaxial growth using Si (001) substrate has been reported by Nagasawa et al. [10]. However, void formation or reduction has not been discussed in these reports.

* Corresponding author. Tel.: +91 3222 283964; fax: +91 3222 255303.
E-mail addresses: agupta_iitkgp@yahoo.com (A. Gupta), joydipdhruba@gmail.com (J.Sengupta), cxj14_holiday@yahoo.com (C.Jacob).

Li and Steckl [5] demonstrated that at the earliest stages of growth, Si diffused out from the substrate non-uniformly, particularly at the periphery of the nuclei. If the nucleation density is low, this results in the formation of irregular trenches, which are then transformed into faceted voids as the growth proceeds. The shape of these voids in a Si (111) substrate is always triangular and in a Si (001) substrate, a square is formed by {111} facets [11,12].

In this paper, we discuss the formation and microstructure of various uncommon shaped voids such as hexagonal, octagonal, etc. along with regular triangular and square voids as observed by optical and atomic force microscopy (AFM) studies in case of 3C-SiC thin film growth on Si by CVD on Si substrates. Truncated voids have been formed in Si (111) and (001) substrates with primary facets on {111} planes. In addition, facets are observed by AFM on the higher energy planes such as {113}, {117}, {112} etc. to form equilibrium void shape. Finally, the reduction of the void density and size has also been investigated by using a two-step growth mechanism.

## 2. Experimental details

Atmospheric pressure chemical vapor deposition of 3C-SiC was carried out on Si (111) and (001) substrates in a hot-wall CVD reactor using a resistance-heated furnace (ELECTRO-HEAT EN345T). Single crystalline Si (111) or Si (001) wafers (p-type) were used as substrates. Before introducing the substrate into the reaction chamber, the native oxide was removed from the substrate surface by dipping in HF solution (5%) for 5 min and then washing in de-ionized (DI) water. After that it was dipped into acetone to eliminate organic contamination, and finally cleaned by DI water. Hydrogen, at a flow rate of 3 slm, was used as the carrier gas and argon, at a flow rate of 3 slm, was used for purging. Propane at a flow rate of 10 to 20 sccm was used for preliminary carbonization just before growth. For actual growth, an organo-metallic single source, hexamethyldisilane (HMDS), at a flow rate of 50 sccm was used. Growth temperature was varied from 1100 °C to 1250 °C. Due to the use of a resistance-heated hot-wall furnace, maximum deposition takes place on the hot wall of the reactor resulting in very low growth rate on the substrates. To grow thicker films, a series of runs was contributed on the same sample sometimes. For this, after each growth, the system was cooled down to room temperature and the reactor was removed from the furnace. The reactor was cleaned and re-growth was carried out on the same sample under the same conditions. The samples were in contact with air during the cleaning process in between two runs in the repeated growth process. The thickness of the film for 2 h growth at 1250 °C is ∼ 0.5 μm.

After growth, the films were characterized using optical microscopy with Nomarski differential interference contrast (LEICA DM LM), XRD (Philips PW1729 X-ray diffractometer using Cu-Kα radiation and θ-2θ geometry) and AFM (Nanonics SPM-100 operated in intermittent contact mode using Nanonics AFM glass probe having force constant 40 N/m and resonant frequency 134.56 KHz) to study the microscopic structure, crystalline nature and defects.

## 3. Results and discussion

### 3.1. X-ray diffraction of SiC films

X-ray diffraction analysis was used in the present work to characterize the crystallinity of the films. Fig. 1a corresponds to diffraction from a film grown on a Si (111) substrate for 2 h at 1250 °C. Similarly, Fig. 1b corresponds to diffraction from the same sample after repeated (two times) growth, which was adopted for growing thicker films at the same temperature. The peak at 28.68° arises from the Si (111) planes due to the CuK$_α$ radiation. In both cases, i.e. Fig. 1a and b, the presence of a peak at ∼ 35.75° indicates that the films are 3C-SiC (111). In case of repeated growth on the same sample, Fig. 1b, peaks at 60.2° and 75.8° appear due to the diffraction from (220) and (222) planes of the grown 3C-SiC film, respectively. In single growth on Si (111) i.e. Fig. 1a, only the presence of 3C-SiC (111) peak proves that the grown film is probably epitaxial. But the appearance of other SiC peaks after repeated growth indicates the polycrystallinity of the film. Therefore it can be concluded that after repeated growth on the same sample, the films become polycrystalline for Si (111). The reason could be that after one run, the sample surface was modified by cooling (stress relief) and exposure to air.

The XRD patterns from the film grown on a Si (001) substrate after first time (2 h at 1250 °C) and after repeated (two times) growth are shown in Fig. 2a and b respectively. The peaks at 69.15° and 41.7° arise due to the diffraction from Si (400) and 3C-SiC (200) planes, respectively. In case of repeated growth, (Fig. 2b), peaks at ∼ 90° appear due to the diffraction from the SiC (400) planes. In Fig. 2a, only the presence of 3C-

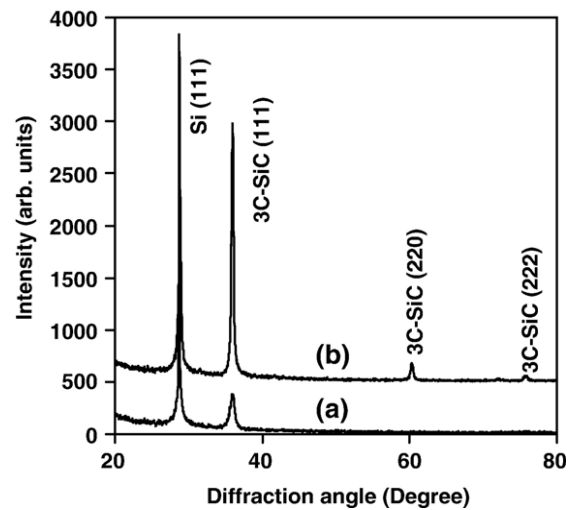

Fig. 1. X-ray diffraction spectra of 3C-SiC films grown on Si (111) substrate (a) for 2 h at 1250 °C and (b) for 4 h at 1250 °C in case of repeated growth.

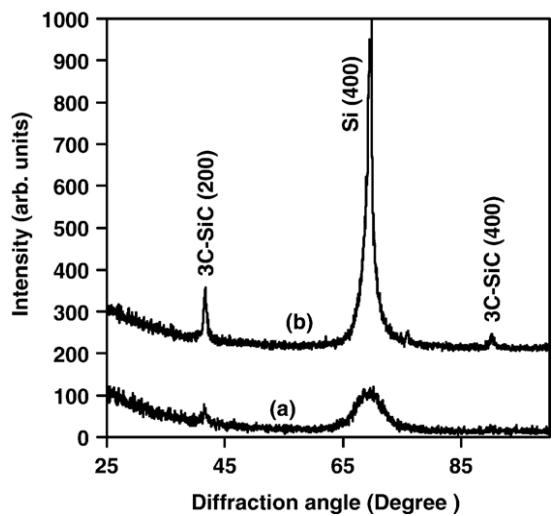

Fig. 2. X-ray diffraction spectra of 3C-SiC films grown on Si (001) substrate (a) for 2 h at 1250 °C and (b) for 4 h at 1250 °C in case of repeated growth.

SiC (200) peak indicates that the grown film is epitaxial. Also, due to the absence of any other intense SiC peak in Fig. 2b, we can say that after two times growth on Si (001) substrate, grown films remain epitaxial. Comparing with Fig. 1b, it can be suggested that better crystallinity has been achieved in case of film grown on Si (001) substrate.

### 3.2. Regular, triangular and square-shaped voids

The optical micrograph of a thin SiC film grown on a Si (111) substrate at 1250 °C for 1 h is shown in Fig. 3. Void formation is thought to be due to Si out-diffusion from the substrate at high temperature (N1100 °C) during carbonization. The triangular shaped objects are the voids. Here the voids are inverted triangular pyramid shapes formed on Si {111} substrate surfaces and bound by {111} planes. The average density of voids is ~ $1.37 \times 10^9$ m$^{-2}$. The perfect triangular shape of the voids was clearly identified using two-dimensional AFM image, as shown in Fig. 4. Here, the void is faceted along

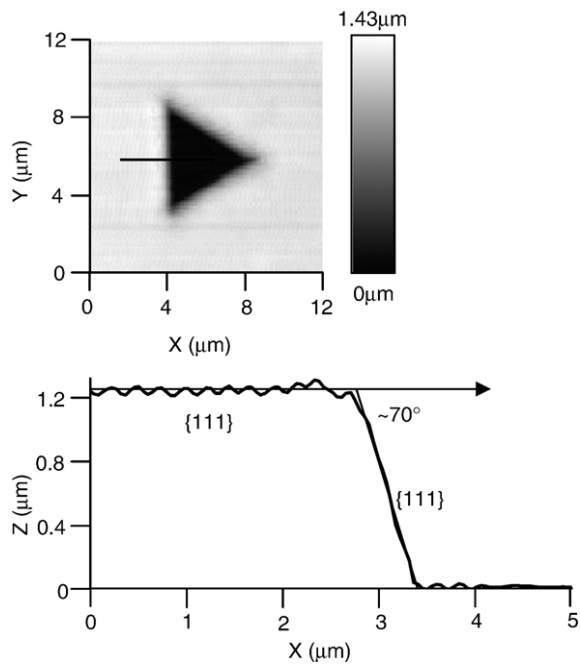

Fig. 4. Two-dimensional AFM image of triangular void in 3C-SiC film grown on Si (111) substrate at 1250 °C for 1 h and cross-sectional image across the void showing facets along {111} planes.

{111} planes because in the cubic closed-packed Si-crystal structure, {111} planes are the lowest surface energy planes (γ (111)=1.23 J/m$^2$) [13].

The cross-section across this triangular void is also shown in Fig. 4 and the depth of the void is ~ 1.4 μm. From this cross-sectional projection, it is clear that the void was not covered by the grown SiC film. The reason behind this was that since a resistance-heated furnace was used, the growth rate was low and therefore, insufficient to seal the voids created during carbonization. Subsequently, Si out-diffusion continues throughout the growth resulting in an increase in the size and depth of the voids. So, voids were created at the Si/SiC interface and extended up to the film surface. From the cross section, it is observed that the angle between the surface and faceting planes is always ∼70°. According to this, it is clear that faceting has taken place along {111} planes, because the angle between {111} planes is ∼70.5°.

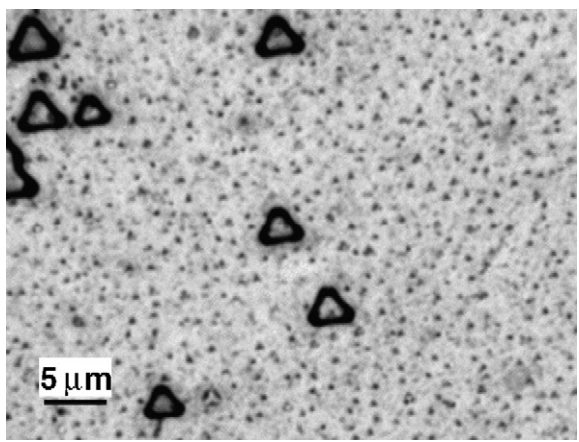

Fig. 3. Optical image of regular triangular voids in 3C-SiC film grown on Si (111) substrate at 1250 °C for 1 h.

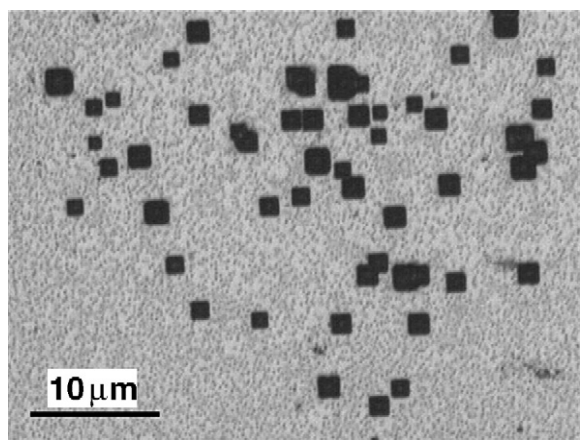

Fig. 5. Optical image of regular square voids in 3C-SiC film grown on Si (001) substrate at 1250 °C for 1 h.

Fig. 5 shows the optical micrograph of a SiC film grown under the same conditions but on Si (001) substrate. The black square/rectangular shaped voids were observed clearly in particular. These voids are hollow inverted pyramids with square or rectangular base planes and faceted along {111} planes. Faceting can be clearly observed from two-dimensional AFM image shown in Fig. 6. Here also, the cross-sections across this void are shown and the depth of the void is ∼1.6 μm. Angle between surface i.e. {001} plane and faceting planes is ∼54.5° which is a strong indication that at the initial stage of formation, faceting takes place along {111} planes because it is the energetically favorable plane.

3.3. Uncommon void formation

Apart from regular triangular voids formed on Si (111), hexagonal voids were also observed regularly depending on the growth conditions. Fig. 7 shows the optical micrograph of the hexagonal voids of larger size along with regular small triangular voids. This sample was grown at 1250 °C for 2 h. Some features are neither an exact triangle nor a perfect hexagon [inset of Fig. 7]. These intermediate features indicate that as growth time increases, truncation of the corners of regular triangular voids occurs with increasing void size and finally perfect large hexagonal voids are obtained. Analyzing the cross sections across a hexagonal void and a triangular void, it was observed that in both the cases faceting takes place along same planes (Fig. 8). For both hexagonal and triangular voids the angle between the surface plane (111) and faceting planes is always ∼70°. From this, it can be concluded that faceting occurred along {111} planes.

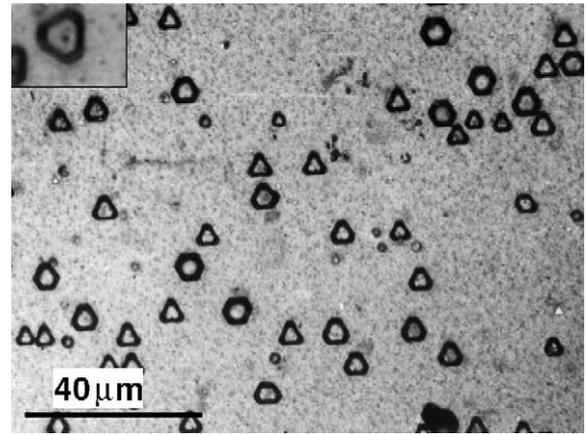

Fig. 7. Optical micrograph of hexagonal voids along with regular triangular voids in 3C-SiC film grown on Si (111) substrate at 1250 °C for 2 h (inset is the intermediate feature showing truncation of triangular void).

Longer growth periods (4 h in two times repeated growth) lead to further truncation of the regular hexagon and void facets along the planes with higher surface energy. Fig. 9 is an optical micrograph showing the truncation of the corners of a hexagonal void. The presence of a small triangular void in the middle of the large hexagon indicates that hexagonal voids are created from triangular void by increasing in size and truncation by facets with higher surface energy. The two-dimensional

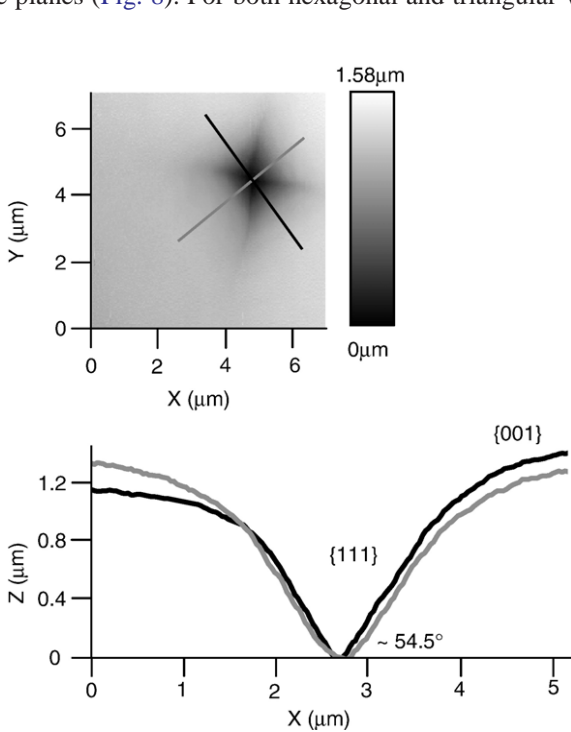

Fig. 6. Two-dimensional AFM image of square void in 3C-SiC film grown on Si (001) substrate at 1250 °C for 1 h on Si (001) and cross-sectional image across the square void showing facets along {111} planes.

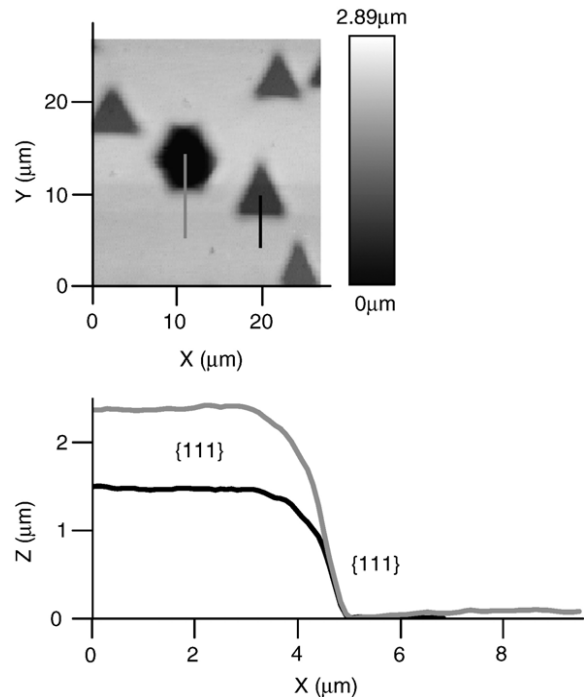

Fig. 8. Two-dimensional AFM image of hexagonal voids in 3C-SiC film grown on Si (111) substrate at 1250 °C for 2 h along with triangular voids and cross-sectional image across the triangular and hexagonal voids showing facets along the same [i.e. {111}] planes.

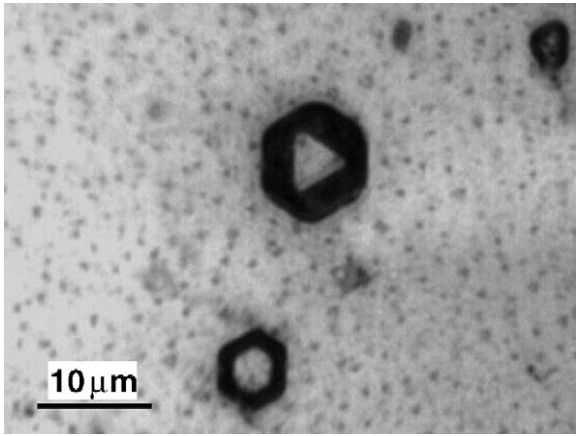

Fig. 9. Optical micrograph showing further truncation of the corners of a hexagonal void in 3C-SiC film grown on Si (111) substrate at 1250 °C for 4 h.

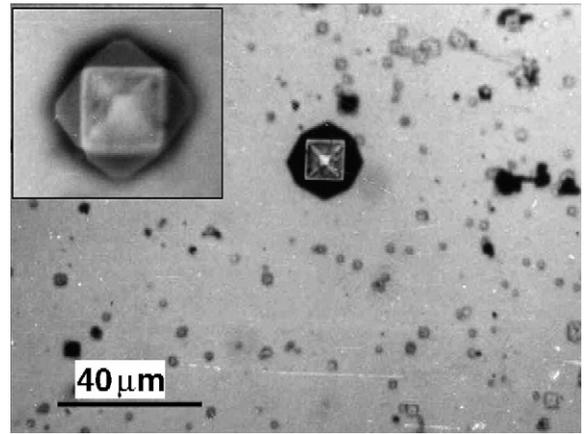

Fig. 11. Optical micrograph of a large octagonal void along with regular square voids in 3C-SiC film grown on Si (001) substrate for 4 h at 1250 °C. Inset is the magnified image of the same void showing the facets.

AFM image (Fig. 10) shows the presence of new planes at the corner of the hexagon. Analyzing the cross section across this feature, it was observed that corner faceting planes are different from the regular {111} planes as the slopes are different. The angle between {111} planes and these corner facets was measured to be ∼ 61°, i.e. these are most probably {112} facets. As {112} surfaces have a higher surface energy than {111} planes so their occurrence could energetically be possible due to the effect of growth kinetics [14].

The uncommon voids are also observed in case of the 3C-SiC films grown on Si (001) substrates as the growth time increases. Fig. 11 is the optical micrograph of the SiC film on Si (001) substrate grown at 1250 °C for 4 h in two times repeated growth, showing a large octagonal void along with regular square voids. Inset is the magnified image of the same void showing the faceting of the void. Fig. 12 shows the two-dimensional AFM image of the octagonal void. In cross section across this octagonal void, instead of faceting along {111} planes, two facets were observed in between the surface and lower {001} planes. These are most likely {117} and {113} planes. As the growth time increases, due to the truncation of the corners of the voids, faceting of the voids on the higher energy planes occur. At high temperature, diffusion around the internal surfaces of the void is high enough, so that all voids can attain their equilibrium shape. It has been reported that, the equilibrium shape for a cubic crystal like Si is a truncated octahedron. It has eight faces in which, the corners are transformed into squares and the triangular faces of the octahedron turned into a regular hexagon. Fig. 13 is the optical image of a large faceted void on a Si (001) substrate. The shape of this faceted void is a truncated octahedron. Thus we can say that all the hexagonal and octagonal voids are close to the equilibrium shape.

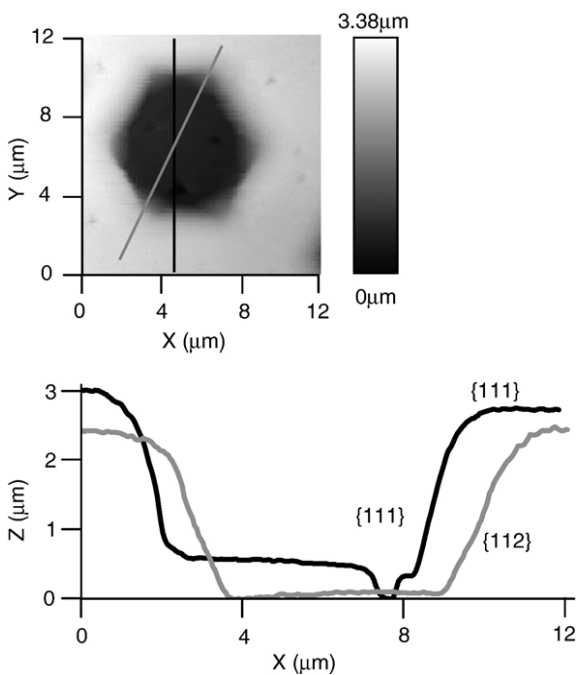

Fig. 10. Two-dimensional AFM image of hexagonal voids in 3C-SiC film grown on Si (111) substrate at 1250 °C for 4 h showing the presence of new planes at the corners of the hexagon and cross-sectional image across the hexagonal voids showing facets.

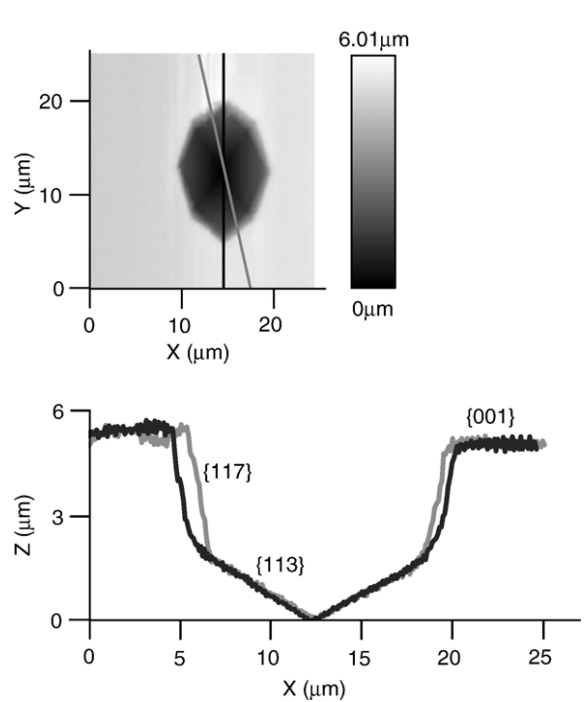

Fig. 12. Two-dimensional AFM image of octagonal voids in 3C-SiC film grown on Si (001) substrate for 4 h at 1250 °C and cross-sectional image across the octagonal void showing facets.

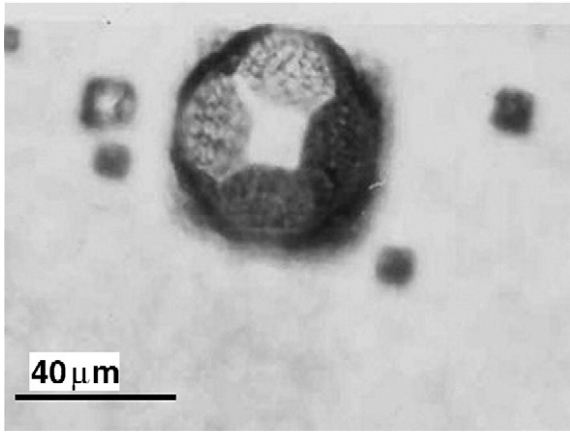

Fig. 13. Optical micrograph of a truncated octahedron in 3C-SiC film grown on Si (001) substrate for 4 h at 1250 °C in case of repeated growth.

3.4. Two-step growth for void reduction

Temperature is one of the main factors affecting the shape, size and density of the voids. Out-diffusion of Si is strongly dependent on temperature. The out-diffusion and the reactivity of Si atoms increase with growth temperature. At low substrate temperatures, out-diffusion of Si atoms from the substrates is negligible. From our experiments, it is clear that SiC nuclei can form at temperatures below 1100 °C and out-diffusion of Si is effective at temperatures above 1100 °C. It was observed that a high density of voids was formed at lower growth temperature (1100 °C-1150 °C) using a preliminary carbonization step prior to the growth. Void free films were grown at temperatures below 1100 °C, but the growth rate was very low. No SiC peak was obtained in XRD analysis at this lower growth temperature (b1100 °C). Finally, for the reduction of void density, a two-step growth process has been used. In this process the initial growth was performed at a lower temperature (1100 °C) using HMDS for 30 min and the second step was done at a higher temperature (1200 °C-1250 °C) without taking out the samples. Our aim was to grow a thin layer of SiC on the Si substrate during the heteroepitaxial deposition step and then grow homoepitaxial single crystalline SiC at higher temperature. The two-step growth method was first introduced by Nishino et al. [2] in the CVD growth of SiC on Si. They introduced a hydrocarbon source to carbonize the silicon surface and to produce a buffer SiC layer to obtain single crystalline 3C-SiC on Si (001). The same group also reported single crystalline 3C-SiC on Si (111) substrates without carbonization [15]. However, in our case, due to the use of a resistance-heated hot-wall furnace, the growth rate is very slow. Therefore, during carbonization by propane, the out-diffusion rate of Si atoms from the surface is higher than the nucleation rate of SiC resulting in a high density of voids. Using preliminary carbonization a very high density of voids was observed even for the low temperature growth (1150 °C). However, by using HMDS in the first stage of our two-step growth process, the nucleation density is higher. Due to the pre-existing Si‑C bonding in HMDS, a low temperature growth is possible in the first stage as compared to the preliminary carbonization. On the other hand, the nucleation density is also higher due to the lower temperature. As a result voids can be reduced using the two-step growth method with HMDS.

X-ray diffraction analysis was used to characterize the crystallinity of the films grown by the two-step method using CuK$_\alpha$ radiation. Fig. 14a corresponds to diffraction from a film grown on a Si (111) substrate using the two-step growth process for a total time of 150 min (1st step: 30 min at 1100 °C; 2nd step: 120 min at 1250 °C). Similarly, Fig. 14b corresponds to diffraction from a sample grown on Si (001) substrate under the same conditions. The presence of only the 3C-SiC (111) peak at 35.85° in Fig. 14a and of the 3C-SiC (200) peak at 41.5° in Fig. 14b, indicates the epitaxial nature of the films.

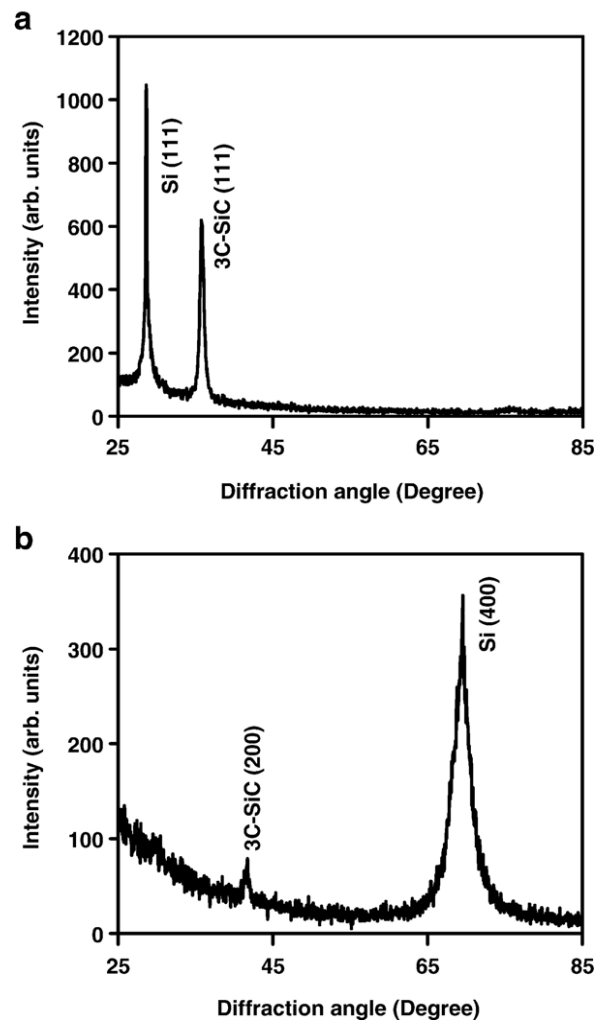

Fig. 14. X-ray diffraction spectra of 3C-SiC films on (a) Si (111) and (b) Si (001) substrates using the two-step growth process process for a total time of 150 min (1st step: 30 min at 1100 °C; 2nd step: 120 min at 1250 °C).

Fig. 15a shows the optical micrograph and Fig. 15b shows a two-dimensional AFM image of a sample grown by two-step growth process on Si (111) substrates with first stage growth at 1100 °C for 30 min and then for 2 h at 1250 °C. One small triangular void is present in optical micrograph but that is not detected by AFM analysis. The reason behind this is the triangular void was sealed in by the film, but due to the optical transparency of the SiC film, it is possible to observe the triangular void at the SiC/Si interface by optical microscopy. In all the cases, no hexagonal voids were observed after a growth of 150 min. Similar effects have been observed in the case of growth on Si (001) substrates using two-stage growth. Fig. 16a shows the optical image of the film grown initially for 30 min at 1100 °C and then for 2 h at 1250 °C on Si (001) substrate. No voids were observed optically. In the AFM image, some islands are observed (Fig. 16b). The out-diffusion rate and the reactivity of Si atoms increase with growth temperature. Thus, it may be concluded that during the first stage of growth at lower temperature, Si out-diffusion was less and on the other hand, due to the use of HMDS and lower growth temperature, the nucleation rate was high. As a result, films with very low void density are obtained due to the rapid coverage of the substrate surface. Occasionally few voids are optically observed but they are completely sealed in by the grown films.

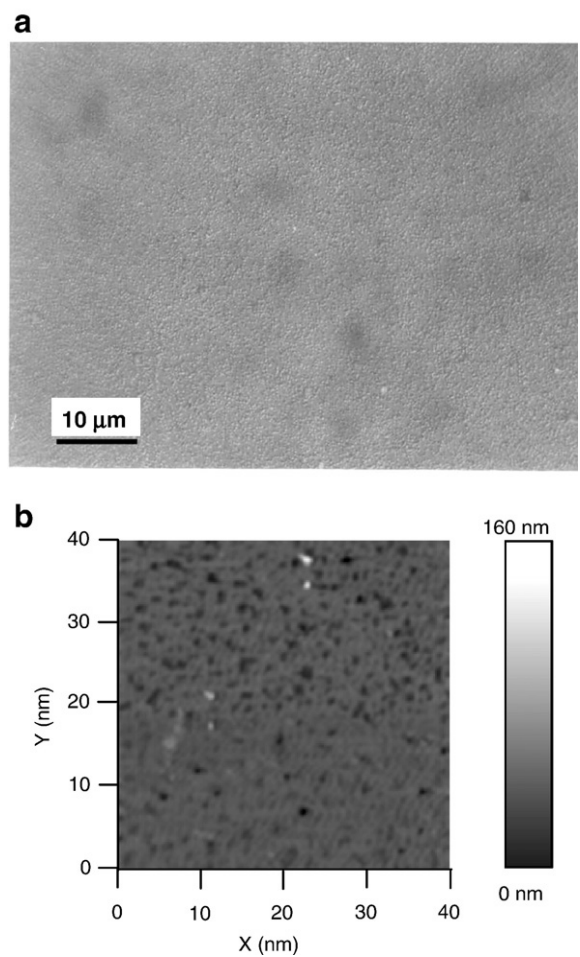

Fig. 16. (a) Optical micrograph of sample grown by two-step growth process for a total time of 150 min (1st step: 30 min at 1100 °C; 2nd step: 120 min at 1250 °C) on Si (001) substrate and (b) Two-dimensional AFM image of the 3C-SiC film grown on Si (001) substrate by two-step growth method.

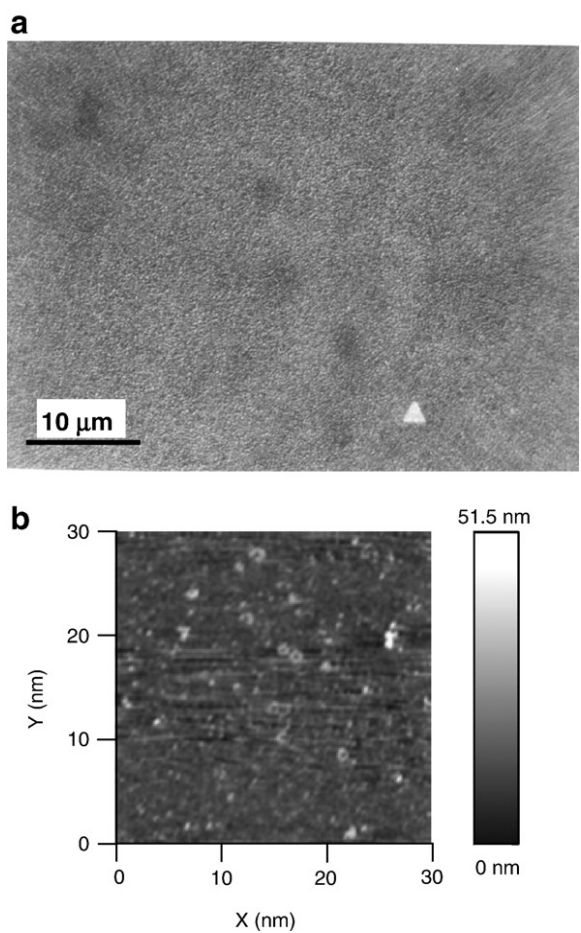

Fig. 15. (a) Optical micrograph of sample grown by two-step growth process for a total time of 150 min (1st step: 30 min at 1100 °C; 2nd step: 120 min at 1250 °C) on Si (111) substrate (b) Two-dimensional AFM image of the 3C-SiC film grown on Si (111) substrate by two-step growth method.

4. Conclusion

In the present study, unusually shaped hexagonal or octagonal voids have been observed along with common triangular and square voids, in heteroepitaxially grown 3C-SiC on Si. The shape of the voids depends on the symmetry of the substrate surface. Primarily, triangular and square shaped voids with {111} facets have been observed in Si (111) and (001) substrates, respectively. With increasing growth time, a more complex shape (e.g. hexagonal and octagonal) of voids is formed by truncation of the regular triangular and square voids. These uncommon voids are faceted not only along {111} planes (lowest surface energy), but also facets with higher surface energy have been observed. The increase of growth time allows the continuation of the diffusion process to approach the equilibrium shape of the voids. The presence of truncated octahedron shows the equilibrium void shape in cubic Si crystals. Finally, the size and density of the voids are drastically reduced using a two-step growth process including low temperature growth with shorter period, followed by high temperature growth. During the first stage of growth at low temperature, the out-diffusion of Si atoms is lower and the deposition rate is higher than by previous carbonization processes resulting in a reduction of voids in the films.

## Acknowledgements

We are grateful for the support provided by DRDO, India. A. Gupta also thanks the CSIR for the award of senior research fellowship.